\documentclass[twocolumn,showpacs,amsmath,amssymb, nobibnotes, aps, prl,showkeys]{revtex4}

\usepackage{graphicx}
\usepackage{dcolumn}
\usepackage{bm}
\usepackage{docs}

\expandafter\ifx\csname package@font\endcsname\relax\else
 \expandafter\expandafter
 \expandafter\usepackage
 \expandafter\expandafter
 \expandafter{\csname package@font\endcsname}%
\fi

\begin{document}

\title{Probing the cosmic ray mass composition in the knee region through TeV secondary particle fluxes from solar surroundings}

\author{Prabir Banik\thanks{Email address: prabir\_banik@yahoo.com}, Biplab Bijay, Samir K. Sarkar and Arunava Bhadra\thanks{Email address: aru\_bhadra@yahoo.com}}

\affiliation{ High Energy $\&$ Cosmic Ray Research Centre, University of North Bengal, Siliguri, West Bengal, India 734013\\}
\begin{abstract}

The possibility of estimating the mass composition of primary cosmic rays above the knee of its energy spectrum through the study of high energy gamma rays, muons and neutrinos produced in the interactions of cosmic rays with the solar ambient matter and radiation has been explored. It is found that the theoretical fluxes of TeV gamma rays, muons and neutrinos from a region around $15^{o}$ of the Sun are sensitive to mass composition of cosmic rays in the PeV energy range. The experimental prospects for detection of such TeV gamma rays/neutrinos by future experiments are discussed.  

\end{abstract}

\maketitle

\section{Introduction}

The sun is known to emit gamma rays during solar flares which are produced in interactions of flare-accelerated particles with solar atmosphere [1]. The prolonged high energy $\gamma$-emissions in solar flares have been detected by the EGRET telescope on board the Compton gamma ray observatory [2] and the Large Area Telescope (LAT) on board the Fermi Gamma-Ray Space Telescope (Fermi) [3] which are believed to be originated from the $\pi^{o}$ decay and thereby an evidence of particle acceleration in solar flare [3]. There was also theoretical predictions that the Sun should radiate gamma rays in the quiescent stage as a result of hadronic interaction of cosmic ray particles with solar atmosphere [4,5] and due to inverse Compton scattering of cosmic-ray electrons on the solar photon halo around the Sun [6]. The EGRET [7] and the Fermi [8] recently observed gamma emission from the quiescent Sun and thereby confirmed the theoretical predictions. Interestingly both the EGRET [7] and the Fermi observations [8] could resolve two components in the quiescent solar radiation - a point like emission from the solar disk which is ascribed to cosmic ray cascades in the solar atmosphere and an extended emission due to the inverse Compton scattering of cosmic ray electrons on solar photons in the heliosphere. The hadronically originated gamma rays should be accompanied by neutrinos of almost the same flux level [9].    

Recently, Andersen and Klein [10] evaluated the fluxes of the high energy gamma rays (photon-pairs) as well as muons and neutrinos from a solid angle within $15^{o}$ around the Sun assuming that the primary cosmic ray particles are all proton at very high energies and concluded that the muon flux so produced might be detectable by next generation air-shower arrays or neutrino detectors [10]. The authors considered only cosmic rays with energies above $10^{16}$ eV, which is roughly the threshold energy for delta resonance production (which subsequently decays into pions and thereby emit gamma rays, muons and neutrinos) in interaction of cosmic ray protons with solar photons. 

An assumption of pure proton as primary cosmic rays is reasonable for estimating low energy gamma rays/neutrino flux as about 90\% of nuclear cosmic rays below few hundred TeV energies are protons. However, when gamma rays/neutrinos in TeV energy range and above from the Sun is concerned, which are likely to be produced in interactions of cosmic rays in the PeV energy range with solar atmosphere, it is important to consider the proper primary composition of cosmic rays above the so called knee energy where the power law spectral index of cosmic ray energy spectrum steepens from $\sim -2.7$ to about $-3.1$ [11]. Note that the primary cosmic rays is studied directly through satellite or balloon borne detectors only up to few hundreds TeV beyond which direct methods become inefficient due to sharp decrease in the flux of primary particles and instead indirect method, through the observation of cosmic ray extensive air shower (EAS), which are cascades of secondary particles produced by interactions of cosmic ray particles with atmospheric nuclei, has to be adopted. Several EAS measurements have been carried out to determine the mass composition of cosmic rays in the PeV energy region and above but the conclusions of different experiments on primary mass composition in the PeV energy region are not unequivocal which is mainly due to the weak mass resolution of the measured EAS observables [12]. Majority of the experiments, however, conclude that the knee represents the energy at which proton component exhibits cut-off (see [12] and references therein) i.e. the knee of the spectrum has been ascribed as the proton knee, which implies that beyond the knee energy, the cosmic ray composition would be heavier, dominated by Fe nuclei. Very recently the KASCADE-GRANDE experiment observed the existence of a Fe-knee around 80 PeV [13] beyond which cosmic rays composition again is dominantly proton. Therefore, the estimation of TeV gamma ray, muon and neutrino fluxes from the Sun due to heavier cosmic ray composition scenario is imperative. Here it is worthwhile to state that the effects of heavier nuclei in cosmic rays have been studied in details for the diffuse galactic gamma rays [14] and neutrino emission [15].   
   
In the present work we would like to analyze the fluxes of high energy gamma rays, muons and neutrinos produced in interaction of high energy cosmic rays with solar radiation and coronal matter as a cosmic ray mass spectrometric technique. For this objective we would extend the previous analysis in several ways. Since the cosmic ray composition above the knee of the cosmic ray energy spectrum is not clearly known, we would consider primary particle can be iron nuclei and we shall consider the whole cosmic ray energy spectrum. However this extension is not simply rerun of Anderson-Klein approach [10] as with the change in nature of primary cosmic rays from proton to Fe, the interaction mechanism becomes complex/ changes as would be elaborated in the following sections. The present work also suggests a way to verify different models for solar coronal matter density through observations of GeV gamma rays/neutrinos from solar corona. Here it is worth mentioning that precise knowledge about solar coronal matter density is an important requirement for resolving cosmic ray mass composition from gamma rays/neutrino observations. Since the cosmic ray composition above the knee of the cosmic ray energy spectrum is not clearly known, we would estimate fluxes of TeV gamma rays, muons and neutrinos produced in interactions of hadronic cosmic rays with solar atmosphere considering different cosmic ray mass composition above the knee and would demonstrate that so produced TeV gamma ray and neutrino flux are sensitive to the primary composition above the knee of the cosmic ray primary energy spectrum. We particularly would estimate the flux of TeV gamma rays/neutrinos from the Sun produced in interaction of primary cosmic rays in interaction with the matter in solar corona as well as in interaction with solar radiation in the vicinity of the Sun considering both proton and Fe primaries beyond the knee. 

The organization of the paper is as the following. In the next section we shall evaluate the TeV gamma rays and neutrino fluxes generated in interaction of cosmic rays with the solar atmosphere. In sub-section 2(a) we shall estimate the TeV gamma rays, muons and neutrino fluxes generated in interaction of cosmic rays with the solar coronal matter. In sub-section 2(b) we would evaluate the TeV gamma rays flux produced in interaction of cosmic rays with solar radiation. We shall discuss our results in section 3 and finally conclude in the section 4.

\section{TeV gamma ray and neutrino fluxes from external region of the Sun}

The cosmic ray particles undergo different interactions with solar atmosphere (matter and radiation) leading to high energy gamma rays/neutrinos. In the corona region the hadronic interaction of cosmic rays with coronal matter dominates over all other interaction processes irrespective of nature of primary (proton or heavier nuclei). Outside the corona, photo production leads to major part of TeV gamma ray/neutrino flux for proton primary. The delta resonance, however, diminishes with nuclear mass and also occurs at relatively higher energies for heavier nuclei. Instead, as we shall see in the later part of the present work, dominant part of TeV gamma ray flux generated by heavier cosmic ray nuclei through photo-disintegration and subsequent de-excitation .

Tll the knee of the cosmic ray energy spectrum i.e. till $\sim 3 \times 10^{15}$ eV we have taken pure proton, pure iron and mixed primary composition following [16] (reasonably consistent with the findings of the direct experiments [17]) whereas above the knee we have considered the following composition scenarios, i) proton primaries till the ankle ($3 \times 10^{18}$ eV) of the cosmic ray spectrum   ii)  Iron primaries up to the ankle iii) the same pre-knee mixed primaries but with rigidity dependent cutoff taking proton cutoff at the knee and iv) Fe primaries up to the second knee ($8 \times 10^{16}$ eV) of the cosmic ray spectrum and proton primary beyond that till the ankle energy. Below the knee the cosmic ray energy spectrum follows a power law [18]

\begin{equation}
 \frac{dn_p}{dE_{p}}(E_{p}< E_{knee})=7.3 \times 10^{19} \frac{eV^{1.7}}{m^2.s.sr} \times E_{p}^{- 2.7}
\end{equation}

as measured by the balloon and satellite based experiments directly. Above the knee the spectral index is $-3.1$ [18] and equating beyond the knee spectrum with below the knee spectrum at the knee position, the absolute intensity of cosmic rays above the knee has been determined. 

\subsection{TeV gamma rays and neutrinos from solar corona}

We have considered the following physical scenario: the high energy (TeV energies and above) cosmic rays traveling towards the Earth interact with the matter in solar corona of the Sun and produce copiously pi-mesons along with other secondary particles. The subsequent decay of $\pi^{o}$ meson gives gamma rays whereas the decay of charged pions give muons and neutrinos. The central part of the solar disk offers a huge thickness of matter to the so-produced gamma rays/muons and the probability of escape of those gamma rays/muons is very very small. In contrast secondary gamma rays/muons produced in solar corona are likely to escape with negligible probability of absorption and might be detected on the earth. Since coronal matter density is very low, the interaction probability of cosmic rays in corona is also small and therefore, the interactions of secondary pions and the leading particles have not been considered. The effects of helio-magnetic field on high energy cosmic rays are negligible and hence ignored.  Here it is worthwhile to mention that Seckel et al [5] considered gamma rays from the whole disk ignoring the coronal part. This is because the low energy gamma rays from the Sun are essentially albedo photons produced in cascades in the solar atmosphere by low energy cosmic rays. The most of the muons are expected to decay enroute and unlikely to reach at Earth. Hence we shall not consider them.   
            
The matter in the coronal part is in the state of plasma, and the particle density is very low, of the order of $10^{15}$ particles/$m^3$ with composition similar to the Sun's interior, mainly ionized hydrogen. Electron density profiles in the heliosphere is inferred from white light brightness measurements of the corona during solar eclipses. The heliosphere is filled by solar wind streams of different velocities and thereby is highly structured; hence the heliosphere radial electron density profile only can be approximated that matches the observations on an average. Several models have been proposed in the literature for radial electron density profiles in the heliosphere but being an approximate description, the model predicted densities differ from each other by some extent. Mann et al [19] obtained a heliospheric density model applicable to a range from the low corona up to five (5) Astronomical units (AU) by solving magneto-hydrostatic equations that include the thermal pressure and the gravitational force of the Sun which is given by

\begin{equation}
N(R_\theta) = N_{S} exp \left[\frac{A}{R_{\odot}}\left(\frac{R_{\odot}}{R_\theta}-1\right)\right]
\end{equation}
where $N_{S}=N(R_{\odot})$, $A=\mu G M_{\odot}/k_{B}T$, $R_{\odot}$ and $M_{\odot}$ are radius and mass of the Sun respectively, $\mu$ is the mean molecular weight, $G$ is the gravitational constant, $k_{B}$ is Boltzmann's constant, $T$ is the temperature and $R_\theta = b/\sin\theta$ is the position of interaction of cosmic ray nuclei where $b$ is the impact parameter of path of cosmic ray nuclei from the centre of sun as shown in figure 1. In the solar corona and the solar wind $\mu$ is about 0.6 [20]. It has been reported that the model agrees very well with different observations for a chosen temperature of $1.0 \times 10^{6} \; K$ such as the electron number density data at low corona extracted from Skylab observations [21], the in-situ particle number density and the particle flux data in the range 0.1 AU to 1 AU from the plasma data of the HELIOS 1 and 2 and IMP satellites [22] (mean deviation around $13\%$) and the particle density data derived from the coronal radio sounding experiment at the ULYSSES spacecraft [23]. We have considered this model for estimating particle flux generated in interaction of cosmic rays with coronal matter. Since coronal density falls sharply with radial distance we have restricted up to three solar radii (equivalently within $1^{o}$ around the Sun). The corona is  electrically neutral, hence we take the ion density in corona to be the same to the electron density. We further assume that the ions in the corona are all protons, so the estimated flux essentially gives the lower bound of the flux. The geometry for interactions of cosmic rays with solar coronal matter or photon is shown in Fig. 1.

\begin{figure}[h]
  \begin{center}
\includegraphics[scale=0.4]{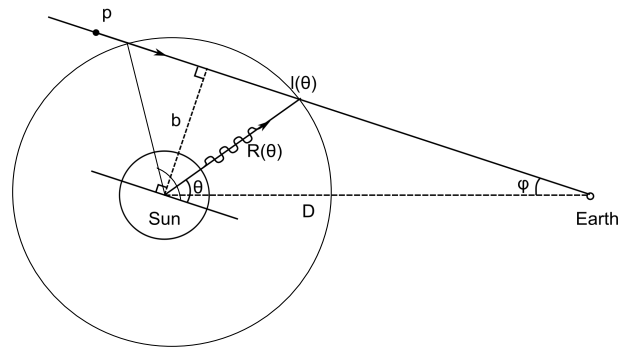}
\end{center}
\caption{Geometry for interactions of cosmic rays with solar photons or coronal matter. }
\end{figure}

For estimation of secondary fluxes it is important to properly take into account the convolution of the corresponding production spectra of secondary particles with the primary CR spectrum, as discussed in [15] for example. When cosmic rays interact with the solar atmospheric nucleons, charged and neutral pions will be produced copiously which will subsequently decay to gamma rays and muons. We have assumed that the inelastic hadronic collisions lead to roughly equal numbers of $\pi^0$, $\pi^+$ and $\pi^-$ mesons. The emissivity of pions, which is assumed to be the same for charged and neutral pions, resulting from interaction of high energy cosmic ray nuclei with coronal matter can be expressed by [24]
 
\begin{eqnarray}
Q_{\pi}^{Ap}(E_{\pi},R_\theta) = c N(R_\theta)\int_{E_{N}^{th}(E_{\pi})}^{E_{N}^{max}}\frac{dn_{A}}{dE_{N}}\frac{d\sigma_{A}}{dE_{\pi}}(E_{\pi},E_{N})dE_{N}
\end{eqnarray}

where $E_{N}^{th}(E_{\pi})$ is the threshold energy per nucleon required to produce a pion with energy $E_{\pi}$ which is determined through kinematic considerations, and $d\sigma_{A}/dE_{\pi}$ is the differential inclusive cross section for the production of a pion with energy $E_{\pi}$ in the lab frame by the stated process. For the inclusive cross section we have used the following model with parametrization of the differential cross section as given by [25,26]

\begin{equation}
\frac{d\sigma_{A}}{dE_{\pi}}(E_{\pi},E_{N}) \simeq \frac{A \sigma_{0}}{E_{N}}F_{\pi}(x,E_{N})
\end{equation}

where $x = E_{\pi}/E_{N}$. The presence of heavier nuclei ($A>1$) in cosmic rays leads to a nuclear enhancement factor A via $ d\sigma_A/dE_{\pi} = A * d\sigma_p/dE_{\pi}$ [14,15]. The inelastic part of the total cross section of p-p interactions ($\sigma_{0}$) is given by [27] 
\begin{eqnarray}
\sigma_{0}(E_{N}) = 34.3+1.88L+0.25L^2 \, mb
\end{eqnarray}
where $L = \ln(E_{N}/TeV)$.

For the energy distribution of secondary pions we used the empirical function as given below [27, 28] that well describes the results obtained with the SIBYLL code by numerical simulations 
 
\begin{eqnarray}
F_{\pi}(x,E_{N}) = 4\beta B_{\pi}x^{\beta-1} \left(\frac{1-x^{\beta}}{1+rx^{\beta}(1-x^{\beta})} \right)^{4} \\ \nonumber
\times \left(\frac{1}{1-x^{\beta}}+\frac{r(1-2x^{\beta})}{1+rx^{\beta}(1-x^{\beta})}  \right) \left(1-  \frac{m_{\pi}}{xE_{N}}\right)^{1/2}\;
\end{eqnarray}

where $B_{\pi} = a + 0.25 $, $\beta = 0.98/\sqrt a $, $a = 3.67+0.83L+0.075L^2$, $r=\frac{2.6}{\sqrt{a}}$ and $L = \ln(E_{N}/TeV)$. The spectra of both charged and neutral pions can be described by the same equation. 

The resulting gamma ray emissivity due to decay of $\pi^0$ mesons thereby can be written as 

\begin{eqnarray}
Q_{\gamma}^{Ap}(E_{\gamma},R_\theta) = 2\int_{E_{\pi}^{min}(E_{\gamma})}^{E_{\pi}^{max}}\frac{Q_{\pi^{0}}^{Ap}(E_{\pi},R_\theta)}{(E_{\pi}^2-m_{\pi}^2)^{1/2}}dE_{\pi}
\end{eqnarray}

where $E_{\pi}^{min}(E_{\gamma}) = E_{\gamma} + m_{\pi}^2/(4E_{\gamma})$ is the minimum energy of a pion required to produce a gamma ray photon of energy $E_{\gamma}$.

\begin{figure}[h]
  \begin{center}
\includegraphics[scale=0.45]{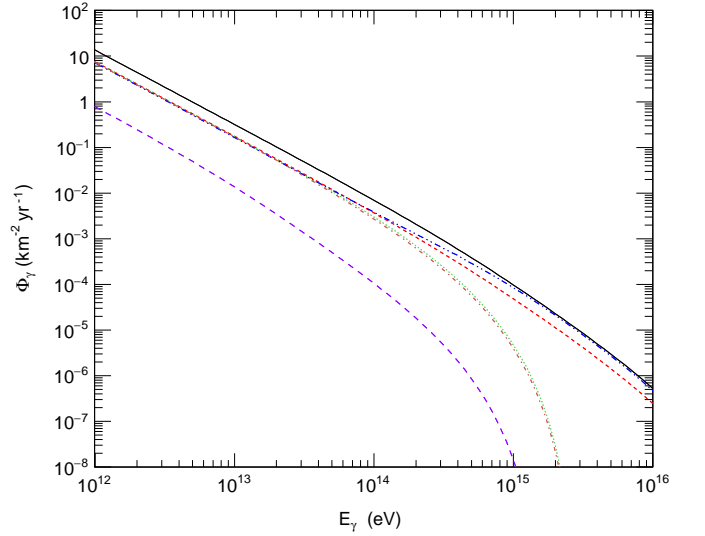}
\end{center}
\caption{Integral energy spectrum of gamma ray photons reaching at earth from the solar corona with observation angle restricted to $1^o$ around the sun. The (black) continuous line corresponds to unchanging proton primary (below and above the knee energy), the (red) small-dashed line describes unchanging mixed cosmic ray composition till the second knee, the (blue) small-dashed-double-dotted line represents mixed cosmic ray composition below the knee that changes to pure proton primary above the knee, the (green) dotted line corresponds to the mixed primaries with rigidity dependent cut-off for all primaries taking proton cut off at the knee , the (pink) dash - dotted line denotes mixed composition below the knee that changes to dominated Fe composition above the knee and the (brown) dashed line corresponds to the iron primary over the whole energy range.}
\end{figure}

The charged pions $\pi^{+}$ and $\pi^{-}$ decay into charged muon and muon neutrinos and anti-neutrinos. The emissivity of such neutrinos can be written as [29]
 
\begin{eqnarray}
Q_{\nu_{\mu}}^{(1)}(E_{\nu_{\mu}},R_\theta) \simeq \frac{m_{\pi}^2}{m_{\pi}^2-m_{\mu}^2}\int_{E_{\pi}^{min}(E_{\nu})}^{^\infty}Q_{\pi^{\pm}}^{Ap}(E_{\pi},R_\theta)\frac{dE_{\pi}}{E_{\pi}}
\end{eqnarray}

For $\left<E_{\nu} \right> \gg m_{\pi} $, the minimum energy of a pion required to produce a neutrino in the stated process is 

\begin{eqnarray}
E_{\pi}^{min}(E_{\nu}) = \frac{m_{\pi}^2}{m_{\pi}^2-m_{\mu}^2}E_{\nu} +\frac{m_{\pi}^4}{m_{\pi}^2-m_{\mu}^2} \frac{1}{4E_{\nu}} \simeq \frac{1}{1-r^2}E_{\nu}
\end{eqnarray}
where $r = m_{\mu}/m_{\pi} $.

On the other hand secondary muons, produced in direct decay of charged pions may subsequently decay $\mu \rightarrow e\nu_\mu \nu_e $ into electrons/positrons and neutrinos.  The contribution of this process to the leptons emissivity is [30]

\begin{eqnarray}
Q_{e}(E_{e},R_\theta) =\frac{m_{\pi}^2}{m_{\pi}^2-m_{\mu}^2}\int_{E_{\mu}^{min}}^{E_{\mu}^{max}}dE_{\mu}\frac{dP}{dE_{e}}P^{'}(\gamma_{\mu},R_\theta) \\ \nonumber
 \times \int_{E_{\mu}}^{E_{\mu}/r^2}\frac{dE_{\pi}}{\beta_{\pi}E_{\pi}} Q_{\pi^{\pm}}^{Ap}(E_{\pi},R_\theta)
\end{eqnarray}

where the three-body decay probability for lepton distribution from a decaying muon is given by [30] 

\begin{eqnarray}
\frac{dP}{dE_{e}} = \frac{8pc}{\beta_{\mu}m_{\mu}^{3}c^{6}}\int du \frac{u(u^2\gamma_{\mu}^2-m_e^2c^4)^{1/2}}{(pc-E_e + u)^2}\left(3-\frac{4\gamma_{\mu}u}{m_{\mu}c^2}\right) \\ \nonumber
\times \left(1-\frac{E_e(E_e - u)}{ p^2c^2}\right)
\end{eqnarray}

where $u = (E_e -\beta_{\mu}pc\cos\alpha)$, p is the electron momentum, the Lorentz factor of muon is $\gamma_\mu = (1-\beta^2)^{-1/2}$ and $\alpha$ is the angle between the direction of produced a lepton and initial direction of decaying muon in lab frame. $P^{'}(\gamma_{\mu},R_\theta)$ is the probability of decay of muon while traveling to the earth and is given by [10]

\begin{eqnarray}
P^{'}(\gamma_{\mu},R_\theta) = 1- \exp\left(-\frac{b/tan\varphi-l_\theta}{c\tau_{\mu}\gamma_{\mu}}\right)
\end{eqnarray}

where $(b/\tan\varphi - l_\theta)$ is the distance from the interaction point to the Earth, $l_\theta = b/\tan\theta$ as shown in figure 1 and $\tau_{\mu} = 2.2 \mu s$ is the muon lifetime at rest.
 
The emissivity of muonic neutrino from the decay of muon can be described by the same function as lepton and hence $Q_{\nu}^{(2)}(E_{\nu},R_\theta) = Q_{e}(E_{e},R_\theta)$ [27]. The total neutrino emissivity due to decay of charged pions by these two processes thus can be written as $Q_{\nu}^{Ap}(E_{\nu},R_\theta) = Q_{\nu}^{(1)}(E_{\nu},R_\theta) +Q_{\nu}^{(2)}(E_{\nu},R_\theta)$.

Using the geometry for interactions of cosmic rays in the solar surroundings as shown in Fig. 1, we have the differential flux of gamma rays and neutrinos reaching at the earth   

\begin{eqnarray}
\frac{d\Phi_{\gamma/\nu}}{dE_{\gamma/\nu}}(E_{\gamma/\nu}) = \int_{R_{\odot}}^{b_{max}} \frac{2\pi bdb}{D \sqrt{(D^{2}-b^{2})}} \\ \nonumber
\int_{\varphi_1}^{\varphi_2}  \frac{bd\theta}{\sin^2\theta}Q_{\gamma/\nu}^{Ap}(E_{\gamma/\nu},R_\theta)
\end{eqnarray}

where $D$ is the distance between the Sun and the Earth, $\varphi_1 = \sin^{-1}(b/3R_{\odot})$ and $\varphi_2 = \pi/2+\cos^{-1}(b/3R_{\odot})$.

The integral flux of gamma rays to be reached the Earth from solar corona as a function of energy is shown in Fig. 2 for different cosmic ray composition scenarios. On the other hand the integral flux of neutrinos to be reached at the Earth from solar corona as a function of energy is shown in Fig. 3 for different cosmic ray composition scenarios.

\begin{figure}[h]
  \begin{center}
\includegraphics[scale=0.45]{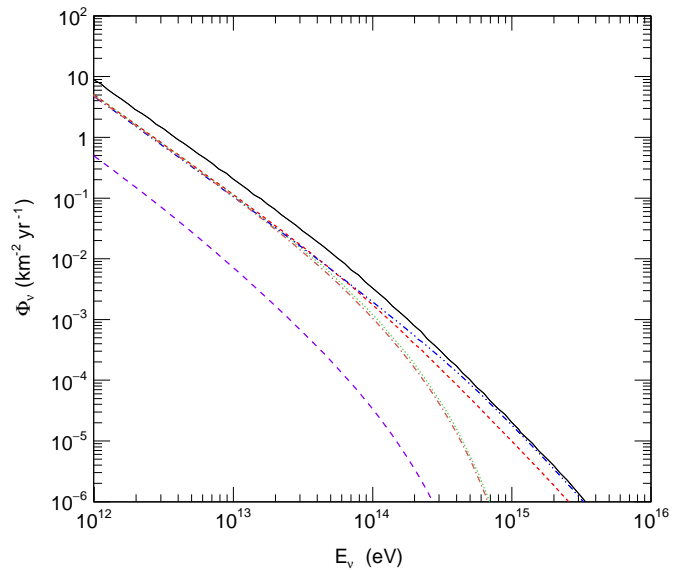}
\end{center}
\caption{Same as the Fig. 2 but for neutrinos (instead of gamma rays).}
\end{figure}

\section{TeV gamma rays produced in cosmic ray interaction with solar radiation}
We shall now explore the impact of primary nuclei on the fluxes of high energy gamma rays which are to be produced in interaction of cosmic rays with solar radiation field. Since the energy of the primary cosmic rays considered here is above few TeV, the deflection of the cosmic rays by the magnetic field of the Sun is negligible and hence is not taken into the account. 

The solar photon flux (in cm$^{-2}$eV$^{-1}$s$^{-1}$) at distance $R(\theta)$ can be estimated from the blackbody spectrum with the temperature of the solar surface $T = 5778$ K and is given by 

\begin{equation}
n_{\gamma}(\epsilon,R_\theta) = \pi \frac{R_{\odot}^2}{R_\theta^2} \frac{2\epsilon^2}{h^{3} c^2}\times \frac{1}{exp(\epsilon/k_bT)-1} .
\end{equation}

where $\epsilon$ is the photon energy, $k_b$ is Boltzmann’s constant, $h$ is Planck's constant and $R_{\odot}$ is the radius of the sun. Since the radiation spreads into a half sphere, a solid angle factor of $\pi$ has been incorporated. 

While traveling in the solar system, a cosmic ray particle encounters a flux of solar radiation. Let a cosmic ray particle with energy $E_{CR}$ and mass $m$ is passing nearby to the sun with an impact parameter $b$ as described in Fig.1. There are two important processes those lead to generation of high energy gamma ray flux as described below.

\subsection{TeV Gamma rays from photodisintegration of cosmic ray iron nuclei}

For heavier nuclei, the photodisintegration is an important process for generating high energy gamma ray flux. In the photodisintegration of high-energy nuclei, the daughter nucleus is typically left in an excited state, which usually immediately emits gamma rays. Here we shall examine the gamma ray flux from the adjacent region of the Sun originated in photodisintegration of cosmic ray nucleus interacting with solar photon flux. Note that the photodisintegration process is employed to explain the recently discovered HEGRA source at the edge of the Cygnus OB2 association [31].

The photo-disintegration rate for a nucleus of atomic number $A$ is given by [32,31] 

\begin{equation}
R_{A}(E_{N},R_\theta) = \frac{1}{2\gamma^2} \int_{0}^{\infty} \frac{n_{\gamma}(\epsilon,R_\theta)d\epsilon}{\epsilon^2} \int_{0}^{2\gamma\epsilon} \sigma_{A}(\epsilon^{\prime})\epsilon^{\prime}d\epsilon^{\prime}.
\end{equation}

where $\sigma_{A}(\epsilon^{\prime})$ is the photdisintegration cross section and $\epsilon^{\prime}$ is the energy of photons in the cosmic ray rest frame.

The nuclear photo-disintegration cross-section is dominated by the giant dipole resonance (GDR) with peaks in the $\gamma$-ray energy range of 10-30 MeV (nuclear rest frame). The process occurs in two-steps which is generally consistent with the experimental data: the nucleus to form a compound state due to photo-absorption, followed by a statistical decay process involving the emission of one or more nucleons from the nucleus [33]. Using Lorentzian model, for a medium and heavy nuclei $A \ge 30$ the total photon absorption cross section can be represented by a Lorentzian  or Breit-Wigner function as given by [26,33]

\begin{equation}
\sigma_{A}(\epsilon^{\prime}) = \sigma_{0}\frac{{\epsilon^{\prime}}^{2}{\Gamma}^{2}}{{{(\epsilon^{\prime}}_{0}}^{2}-{{\epsilon^{\prime}}^{2}})^{2}+{\epsilon^{\prime}}^{2}{\Gamma}^{2}}
\end{equation}

where ${{\epsilon^{\prime}}_{0}}$ is the position of the GDR, $\Gamma$ is the width of the resonance, and $\sigma_{0}$ is the normalization constant.

With the single pole of narrow-width approximation, the cross section can be safely approximated as [26]

\begin{equation}
\sigma_{A}(\epsilon^{\prime}) = \pi\sigma_{0}\frac{\Gamma}{2}\delta(\epsilon^{\prime} -\epsilon^{\prime}_{0} )
\end{equation}

where $\sigma_{0}/A = 1.45\times 10^{-27} $ cm$^{2}$, $\Gamma = 8$ MeV, and $\epsilon^{\prime}_{0} = 42.65A^{-0.21}$ ($0.925A^{2.433}$) MeV for A$ > 4$ (A$ \le 4$). Putting Eq.17 in Eq.15, we obtain [22]

\begin{align}
R_{A}(E_{N},R_\theta) &\simeq \frac{\pi\sigma_{0}\epsilon^{\prime}_{0}\Gamma}{4\gamma^2}\int_{0}^{\infty}\frac{d\epsilon}{\epsilon^2}n_{\gamma}(\epsilon,R_\theta)\Theta(2\gamma\epsilon -\epsilon^{\prime}_{0} )  \nonumber \\  
  &= \frac{\pi\sigma_{0}\epsilon^{\prime}_{0}\Gamma}{4\gamma^2}\int_{\epsilon^{\prime}_{0}/2\gamma}^{\infty}\frac{d\epsilon}{\epsilon^2}n_{\gamma}(\epsilon,R_\theta)
\end{align}

Ultra high energy(UHE) cosmic ray heavy nuclei with Lorentz factor $\gamma =  E_{CR}/Am_{p}$ undergo photo-disintegration when they interact with the solar photon radiation field and release a number of protons and neutrons in the process. The rate of photodisintegration of cosmic ray iron nuclei by the solar radiation at $R_{\theta} =  2R_{\odot}/$ as a function of cosmic ray energy is shown in figure 4. It is seen that the photodisintegration rate increases sharply till about $2 \times 10^{17}$ eV, thereafter it becomes roughly constant.    

\begin{figure}[h]
  \begin{center}
\includegraphics[scale=0.45]{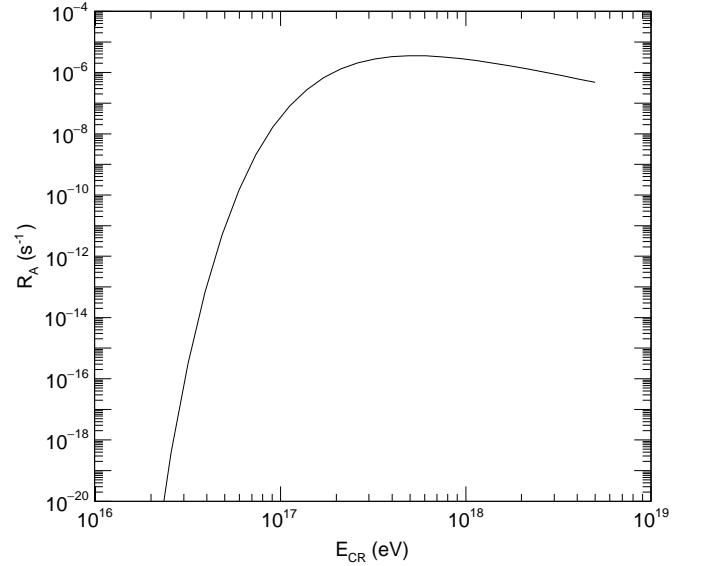}
\end{center}
\caption{The rate of photodisintegration of cosmic ray iron nuclei by the solar photons at $R_{\theta} =  2R_{\odot}/$ as a function of cosmic ray energy.}
\end{figure} 

Approximating the $\gamma$-ray spectrum as being mono-chromatic with energy equal to its average value ($\bar{E}^{\prime}_{\gamma,A}$), the emissivity of gamma ray photon due to de-excitation of photo-disintegrated nuclei can be expressed as  [26]

\begin{equation}
Q_{\gamma}^{dis}(E_{\gamma},R_\theta) = \frac{\bar{n}_A m_{N}}{2 \bar{E}^{\prime}_{\gamma,A}}  \int_{\frac{m_NE_{\gamma}}{2 \bar{E}^{\prime}_{\gamma,A}}} \frac{dn_A}{dE_N}(E_N)R_{A}(E_{N},R_\theta) \frac{dE_N}{E_{N}}
\end{equation}

where $\bar{E}^{\prime}_{\gamma,A} $ is the average energy of the emitted gamma ray photon, $\bar{n}_A$ represents the mean $\gamma$-ray multiplicity for a nucleus with mass number $A$ and $m_N $ is the rest mass of each nucleon. Here we have used $\bar{E}^{\prime}_{\gamma,56} = 2$ MeV and $\bar{n}_{56} = 2$ for iron nuclei [26]. The photon flux at energy $E_{\gamma}$ is produced from cosmic rays with (per nucleon) energy $E_{N}^{min}= E_{CR}/A=  \frac{m_N E_{\gamma}}{2 \bar{E}^{\prime}_{\gamma,A}}$ and above. The upper limit of the integral in Eq.(19) is taken as the maximum cosmic ray energy considered.

The flux of gamma ray photons at earth follow from photodisintegration process is therefore given by 
\begin{equation}
\frac{d\Phi_{\gamma}}{dE_{\gamma}}(E_{\gamma}) = \int_{b_{min}}^{b_{max}} \frac{2\pi bdb}{D \sqrt{(D^{2}-b^{2})}} \int_{\varphi}^{\pi}  \frac{b d\theta}{\sin^2\theta}Q_{\gamma}^{dis}(E_{\gamma},R_\theta)
\end{equation}
where $b_{min} = D.\sin2^{o}$ and $b_{max} = D.\sin15^{o}$ which are effective impact parameters for a observation of $2^{o}$ to $15^{o}$ solid angle about the sun as seen from earth. 

If we assume that all cosmic ray particles are Iron between the cosmic ray knee to the ankle, the integral flux of gamma ray photons above 10 GeV is found 0.026 particles/(km$^2.$ yr). Instead if we restrict primary energy up to the second knee only the integral flux of the gamma ray photons above 10 GeV is 6.3$\times10^{-5}$ particles/(km$^2.$ yr) which implies that the dominant contribution comes from cosmic rays (Fe) above the second knee to the ankle energy as suggestive from the Fig. 4. The resultant differential spectrum of gamma ray reaching the Earth is shown in Fig 5. The figure implies that the photo-disintegration process dominates after the second knee of the cosmic ray spectrum.

\begin{figure}[h]
  \begin{center}
\includegraphics[scale=0.45]{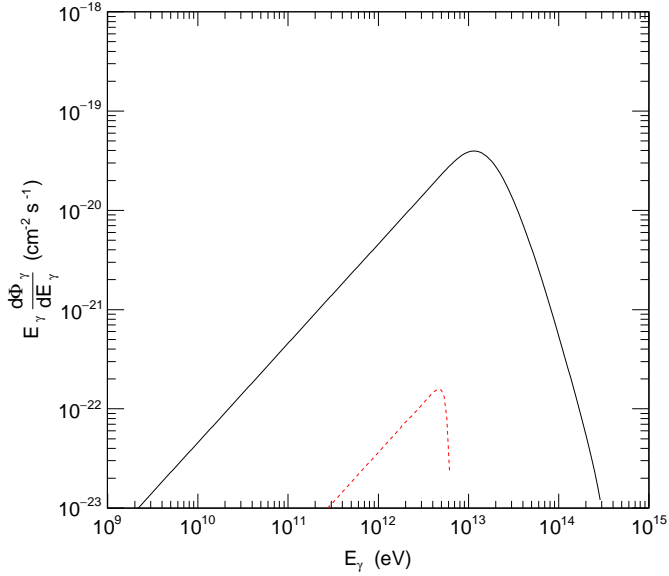}
\end{center}
\caption{Differential energy spectrum of photons reaching at the Earth from 2$^o$ to 15$^o$ solid angle area around the Sun. The black continuous line represents the photon flux considering mixed composition of cosmic rays below the knee and iron above the knee till the ankle energy whereas the red dashed line describes the same (to that of the black continuous line) but restricting the upper energy up to the second knee only.}
\end{figure} 

\subsection{TeV Gamma rays from photo-pion interaction of cosmic ray nuclei:}

The photo-pion interaction is another important process for generating high energy gamma ray flux. As UHE cosmic ray nuclei propagate through vicinity of the solar system, they produce pions through photo-pion interactions with the radiated solar photons at a rate [34,33]

\begin{equation}
G_{A}(E_{N},R_\theta) = \frac{1}{2\gamma^2} \int_{\epsilon_{th}^{\prime}/2\gamma}^{\infty} \frac{n_{\gamma}(\epsilon,R_\theta)d\epsilon}{\epsilon^2} \int_{\epsilon_{th}^{\prime}}^{2\gamma\epsilon} \sigma_{A}(\epsilon^{\prime})\epsilon^{\prime}k(\epsilon^{\prime})d\epsilon^{\prime}.
\end{equation}

where $k$ is the inelasticity coefficient and $\epsilon_{th}^{\prime} = 0.15$ GeV is the threshold energy of solar photon in centre of mass frame. The main contribution to the second integral in Eq. 21 is from the photon energies $\epsilon^{\prime} \sim \epsilon^{\prime}_{0} = 0.3$ GeV, where the cross section peaks due to the $\Delta$ resonance.

Approximating the integral by the contribution from the resonance one obtains [34]

\begin{equation}
G_{A}(E_{N},R_\theta) = \frac{k(\epsilon^{\prime}_{0})\sigma^{A}_{0}\epsilon^{\prime}_{0}\Delta\epsilon^{\prime}}{2\gamma^2} \int_{\epsilon_{0}^{\prime}/2\gamma}^{\infty} \frac{n_{\gamma}(\epsilon,R_\theta)d\epsilon}{\epsilon^2} 
\end{equation}

where $\sigma^{A}_{0}/A \simeq 5\times 10^{-28} $ cm$^{2}$ and $k(\epsilon^{\prime}_{0}) \simeq 0.2$ are the values of $\sigma$ and $k$ at $\epsilon^{\prime} = \epsilon^{\prime}_{0}$, and $\Delta\epsilon^{\prime} \simeq 0.2$ GeV is the peak width of the resonance.

By replacing $\bar{E}^{\prime}_{\gamma} = km_{N}/2$, $\bar{n} = 2$ and $R_{A}(E_{N},R_\theta) \rightarrow G_{A}(E_{N},R_\theta)$ in Eq.($19$), one gets the emissivity of gamma rays as [35]

\begin{equation}
Q_{\gamma}(E_{\gamma},R_\theta) = \frac{2}{k}\int_{E_{\gamma}/k}\frac{dn_A}{dE_N}(E_N)G_{A}(E_{N},R_\theta)\frac{dE_N}{E_{N}}
\end{equation}

Substituting the above equation in the Eq. (20) in place of $Q_{\gamma}^{dis}(E_{\gamma},R_\theta)$, we calculate the flux of gamma rays at earth due to photopion interaction.

Assuming all cosmic ray particles are protons, the total flux of gamma ray photons is found about $4.2\times10^{-4}$ particles/(km$^2$·yr) in a solid angle range $2^o$ to $15^o$ around the Sun. Instead if all cosmic ray particles are Iron in the same energy range then the flux of photons will be about $3.1\times10^{-6}$ particles/(km$^2$·yr) from the same region around the Sun. The variation of the differential energy spectra of the created photons reaching at earth for pure proton and pure Fe primaries are shown in Fig. 6.

\begin{figure}[h]
  \begin{center}
\includegraphics[scale=0.45]{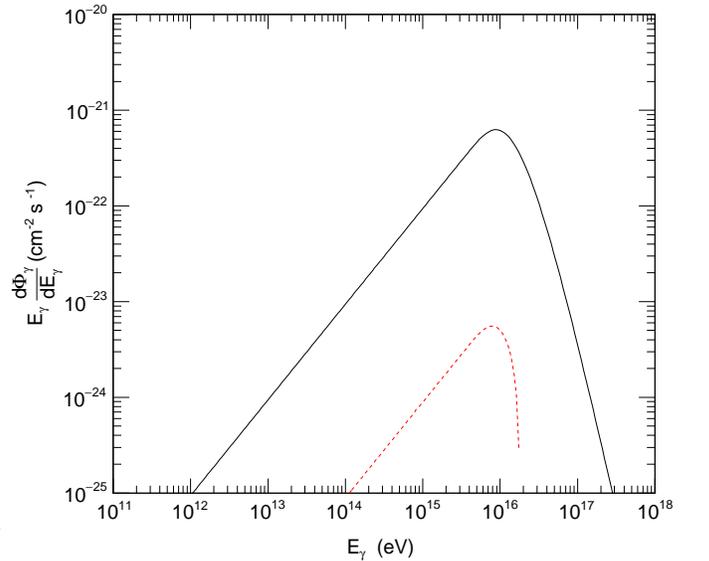}
\end{center}
\caption{Differential Energy spectrum of photons in a 2$^o$ to 15$^o$ area around the Sun where black continuous line and red dashed line represents the gamma ray flux considering proton and iron as a primary CR spectrum respectively.}
\end{figure}

\section{Discussion and Conclusion }

We have analyzed the production of TeV gamma rays/neutrinos in the external part of the Sun through interaction of high energy cosmic rays with coronal matter and solar radiation field considering different cosmic ray mass composition above the cosmic ray energy spectrum and estimate the corresponding fluxes at the Earth. It is found that TeV gamma ray and neutrino fluxes from the solar corona and a region $15^{o}$ observation angle about the sun are sensitive to the primary composition of cosmic rays above the knee of the cosmic ray energy spectrum and thereby can be utilized, at least in principle, to estimate the cosmic ray composition above the knee energy.

In the case of TeV gamma rays/neutrinos from the solar corona, the estimated fluxes remain uncertain to some extent due to lack of precise knowledge of matter density profile in solar corona. For instance, the radial density profile model of Mann et al [20] considered here has a difference of around $15\%$ from the well-known fourfold Newkirk model [36] that was developed in accordance with the observations of white light scattering in the corona during a solar minimum period. The primary composition of cosmic rays is well known up to few hundred TeV from direct measurement. Since the lower energy part of the gamma ray spectrum is not affected by cosmic rays of PeV energies or higher, uncertainty in the flux of gamma rays/neutrinos in the GeV energies from solar corona comes solely from the uncertain matter density profile of solar corona. Hence precise measurement of GeV gamma ray flux in principle should discriminate the different models for radial matter density in solar corona. Once the correct model for density profile is identified from the GeV gamma ray observations, the observed TeV and PeV gamma ray fluxes can be utilized to discriminate primary cosmic ray composition scenario above the knee of the cosmic ray energy spectrum.

\begin{figure}[h]
  \begin{center}
  \includegraphics[width = 0.5\textwidth,height = 0.45\textwidth,angle=0]{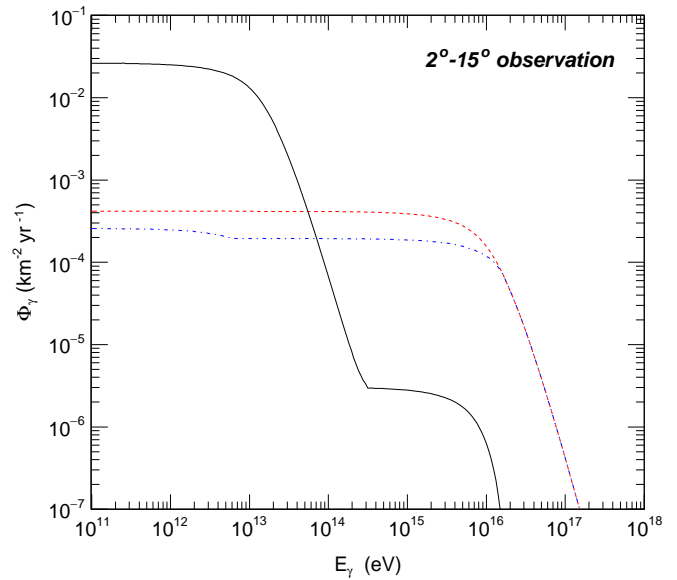}
\end{center}
  \caption{Integral energy spectrum of photons reaching the Earth from a solid angle $2^{o}$ to $15^{o}$ around the Sun. The (red) dashed line, the (black) continuous line and the blue dotted line respectively describe the photon flux when the cosmic rays are (a) pure proton, (b) proton up to the knee and iron above the knee energy and (c) proton up to the knee, iron above the knee up to the second knee and again proton above the second knee.}
\end{figure}

When interactions of energetic cosmic rays with solar photons are considered, it is found that if cosmic rays are Fe nuclei above the second knee energy, the TeV gamma ray flux would be substantially (nearly two order) higher than that due to proton dominated composition at the same energy region. Because of the lower threshold energy, the gamma ray production by energetic cosmic ray Fe nuclei interacting with solar radiation field is dominated by photodisintegration process  whereas the photoproduction process leads the gamma ray production in interaction of cosmic ray protons with the solar photons. The gamma ray, flux through photoproduction of cosmic ray Fe nuclei is much less than those produced by photodisintegration process as well as in photoproduction of cosmic ray protons. Since photodisintegration process does not lead to neutrino and muon, the neutrino and muon fluxes (through photoproduction) for cosmic ray Fe nuclei are about two order less than those due to cosmic ray protons. A point to be noted that the photodisintegration of cosmic ray nucleus in interaction with the solar radiation may give rise to development of two separate air showers almost simultaneously at some spatial separation initiated by two photodisintegrated daughter nuclei [37,38] which might be observed by the ongoing/future cosmic ray air shower arrays if cosmic rays at those energies are indeed heavy nuclei.

The question is that whether TeV gamma rays/neutrinos produced in interaction of high energy cosmic rays with solar coronal matter/ solar radiation can be observed or not experimentally. It appears that the chances of observation is better from solar corona than outside the corona. It is found that around $0.5$ TeV and slightly below that there should be few tens of events per year in a square KM gamma ray observatory. The proposed Cerenkov Telescope Array (CTA) is supposed to have such large collection area but being a Cerenkov imaging telescope, it can not see the Sun. Instead water Cerenkov experiments such as MILAGRO [39] ot HAWK [40] can detect gamma rays from and around the Sun but their collection area is much smaller. Only a square kilometer extension of HAWK kind of experiment should able to detect cosmic ray induced TeV gamma rays from solar corona and thereby may address the mass composition of cosmic rays above the knee. 

The TeV gamma ray flux produced in interaction of high energy cosmic rays with solar photons around the Sun is a mass sensitive observable. For Fe primary above the second knee the so produced Tev Gamma ray flux is about two order higher than that due to proton primary. However, such TeV gamma rays is unlikely to be detected in near future experimentally even if cosmic rays are Fe above the second knee of the spectrum. The integral so produced TeV gamma ray flux from a region $2^{o}$ to $15^{o}$ around the Sun is shown in figure 7. It appears from the figure that to observe at least one event per year when the cosmic rays are Fe above the second  knee, the detector area need to be nearly 10 square kilometer. 

Andersen and Klein pointed out that the muon flux from solar surroundings produced in photoproduction of cosmic ray protons should be detectable by the future experiments [10]. The present analysis suggests that the muon flux due to Fe dominated cosmic ray composition around 100 PeV energy will be much smaller than those due to proton dominated composition at such high energies. This implies that if future experiments really can see the appropriate muon flux from solar surroundings, the primary composition of cosmic rays in the 100 PeV range can be conclusively inferred as proton dominated.

\section*{Acknowledgments}
The authors would like to thank an anonymous reviewer for insightful comments and very useful suggestions that helped us to improve and correct the manuscript.

\end{document}